# Dual-Topology Hamiltonian-Replica-Exchange Overlap Histogramming Method to Calculate Relative Free Energy Difference in Rough Energy Landscape

Donghong Min[3][1], Hongzhi Li[3][1], Guohui Li[1], Ryan Bitetti-Putzer[3], Wei Yang[123]*

1. *Department of Chemistry and Biochemistry, Florida State University, Tallahassee, FL, 32306*
2. *Institute of Molecular Biophysics, Florida State University, Tallahassee, FL, 32306*
3. *School of Computational Science, Florida State University, Tallahassee, FL, 32306*

*A novel overlap histogramming method based on Dual-Topology Hamiltonian-Replica-Exchange simulation technique is presented to efficiently calculate relative free energy difference in rough energy landscape, in which multiple conformers coexist and are separated by large energy barriers. The proposed method is based on the realization that both DT-HREM exchange efficiency and confidence of free energy determination in overlap histogramming method depend on the same criteria: neighboring states' energy derivative distribution overlap. In this paper, we demonstrate this new methodology by calculating free energy difference between amino acids: Leucine and Asparagine, which is an identified chanlleging system for free energy simulations.*

PACS Codes:

Calculating relative free energy difference has been one of major subjects in computational biophysics. It can help us to quantitatively understand solvation, drug binding, and protein-protein interaction etc. at atomistic level [1]. Accuracy, efficiency and robustness are the most concerned issues in the development of free energy simulation techniques [1]. Commonly, free energy simulations are performed in the framework of equilibrium thermodynamics. For instance, in "alchemical" free energy perturbation (FEP) [2] and thermodynamics integration (TI) [3] techniques, energy potentials of two states are connected by scaling parameter, $\lambda$ (usually linearly). Slowly switching $\lambda$ (including long time relaxation at each fixed $\lambda$ value) allows reversible transformations between two states, as shown below:

$$U = (1-\lambda)U_A + \lambda U_B \quad . \tag{1}$$

Traditional FEP and TI methods have been proved to be valuable and served computational communities a great deal [1]. However, their shortcomings in the issues mentioned above are becoming obvious, especially after exhaustive applications [1,4]. Recently, improvement in free energy simulation techniques has been mostly emphasized in efficiency issues [4,5]. It is especially motivated by the proposals of non-equilibrium free energy simulation theories, such as Jarzynsky formulism [6] and overlap histogramming method [7] (or sometimes called "fluctuation theorem" or "Crooks theorem" [8]).

In non-equilibrium free energy simulations, scaling parameter λ can be switched in fast manner, from ensembles of starting state A to targeting state B in the forward direction or from ensembles of targeting state B to starting state A in the reverse direction. Following the tradition [7], we can label the work ($W$) distribution in the forward direction as $f(W)$ and backward direction as $g(W)$. These two distributions can be utilized to obtain relative free energy difference $\Delta A$ through the following relationship:

$$f(W)e^{\Delta A/k_B T} = g(W)e^{W/k_B T} ; \tag{2}$$

This relationship is called "overlap histogramming". In this method, determination of the intersection point between two distributions allows obtaining relative free energy difference of two states, because when $f(W) = g(W)$, $W = \Delta A$. Confidence of determining the intersection point relies on how well these two distributions overlap. Usually, biomolecular systems have multiple conformers separated by possibly large energy barrier, which can prevent an effective exploration of the configurational space. So conformational sampling is another issue to be concerned about, in order to improve free energy simulation efficiency and accuracy, especially in rough energy landscape [9].

Here, we introduce an efficient method, which can solve these two problems simultaneously in a robust manner. In the present method, scaling parameter λ serves two purposes both as effective temperature label (for the purpose of enhancing sampling) and potential energy scaling parameter (to improve the confidence in determining the intersection point). The present development is based on Hamiltonian Replica Exchange Method (HREM) [10]; it is a variant of more often used Temperature Replica Exchange Method (TREM) [11] by replacing the random walk in temperature with one through an ensemble of models with scaled (or partially scaled) energy function: $U = U_E + \lambda U_S$. In HREM, sampling in the portion where potential is scaled, can be very efficient, due to the reason that rate-limiting energy barrier crossing can be facilitated by exchanging structures with simulations on low $\lambda$ energy functions. Here, $\lambda$ can be considered as the label of local "effective temperature" [12]. HREM method is particularly useful in simulating local events in biomolecular systems, such as ligand binding or local conformational change, because of its flexibility in treating local potentials without destructing global structural integrity. This treatment can also help to improve exchange rate [10]. In our HREM simulation, switching potential function for free energy simulations shown in Eq. 1 can be rewritten as

$$U = (1 - \lambda)U_S^A(\overrightarrow{X}) + \lambda U_S^B(\overrightarrow{X'}) + U_E, \qquad (3)$$

in which part of system incorporating the unique portion of starting state is scaled by (1-$\lambda$) and the other one incorporating the unique portion of targeting state is scaled by $\lambda$. Scaled portions have independent coordinates $\overrightarrow{X}$ and $\overrightarrow{X'}$, and so at two end points, original starting state and targeting state energy functions can be obtained. Different from regular HREM, the present method, named by us, "Dual-Topology HREM" (DT-HREM), has two parts representing two end states, scaled in the reverse directions, independently by $\lambda$ and (1-$\lambda$). Here, $\lambda$ and (1-$\lambda$) become labels of local effective temperatures besides original roles as potential scaling parameters in alchemical free energy simulations. $N$ noninteracting copies of the molecule simulated on Dual-Topology potentials with a series of $\lambda$ values ($\lambda_1 = 0 < \lambda_2 < \lambda_3 \ldots < \lambda_N = 1$), shown in Eq. 3, are exchanged according to

$$w(C^{old} \to C^{new}) = \min(1, \exp\{-\beta[U_i^j + U_j^i - U_i^i - U_j^j]\}) = \min(1, \exp\{-\beta$$
$$[(1-\lambda_j)U_S^A(\overrightarrow{X_i}) + \lambda_j U_S^B(\overrightarrow{X'_i}) + U_E(\overrightarrow{C_i}) + (1-\lambda_i)U_S^A(\overrightarrow{X_j}) + \lambda_i U_S^B(\overrightarrow{X'_j}) +$$
$$U_E(\overrightarrow{C_j}) - (1-\lambda_i)U_S^A(\overrightarrow{X_i}) - \lambda_i U_S^B(\overrightarrow{X'_i}) - U_E(\overrightarrow{C_i}) - (1-\lambda_j)U_S^A(\overrightarrow{X_j}) - \lambda_j U_S^B$$
$$(\overrightarrow{X'_j}) - U_E(\overrightarrow{C_j})]\}) = \min(1, \exp\{-\beta\Delta\lambda\Delta U_S^{AB}\})$$

(4)

Here, $\Delta\lambda = \lambda_j - \lambda_i$ and $\Delta U_S^{AB} = [(U_S^B(\overrightarrow{X'_i}) - (U_S^A(\overrightarrow{X_i})] - [(U_S^B(\overrightarrow{X'_j}) - (U_S^A(\overrightarrow{X_j})]]$; $\Delta U_S^{AB}$

is also the difference of energy derivatives $\left.\dfrac{\partial U}{\partial \lambda}\right|_{\lambda=\lambda_i}$ and $\left.\dfrac{\partial U}{\partial \lambda}\right|_{\lambda=\lambda_j}$. So exchange frequency

is determined by the overlap of energy derivative distributions at ensembles $\Gamma(\lambda_i)$ and

$\Gamma(\lambda_j)$. Because of this exchange move, all the energy landscapes in $\overrightarrow{X}$ and $\overrightarrow{X'}$ portion

in an ensemble models with various λ values can be explored efficiently, provided that sampling of $\overrightarrow{X}$ and $\overrightarrow{X'}$ is aided independently by high-λ end and low-λ end simulations respectively. This is due to the reason that at each of two ends of scaling parameters,

energy barriers for one of states are scaled down to help all the copies with λ values near

the other ends "tunnels" through energy barriers. And so at each λ simulation, ensemble

$\Gamma(\lambda)$ for its corresponding potential (Eq. 3) can be generated efficiently.

For two ensembles generated in DT-HREM $\Gamma(\lambda_i)$ and $\Gamma(\lambda_j)$, free energy difference $\Delta A_{ij}$ in between them can be calculated by applying Eq. 2. Considering instantaneous move between these two states, forward work and backward work should be energy difference $\Delta U_{ij}$ (between λ=$\lambda_j$ and λ=$\lambda_i$ potentials), on the structures in $\Gamma(\lambda_i)$

(forward) and $\Gamma(\lambda_j)$ (backward) ensembles respectively. Since $\Delta U$ is equal to $\dfrac{\partial U}{\partial \lambda}$ times

$\Delta\lambda_{ij}$, confidence of determining intersection of two work ($\dfrac{\partial U}{\partial \lambda}\Delta\lambda_{ij}$) distributions should

depend on the overlap of energy derivative distributions as replica exchange frequency

does. So in the present "Dual-Topology Hamiltonian-Replica-Exchange Overlap

Histogramming Method", configurational space sampling problem and work distribution function overlap problem can be solved simultaneously, because they are regulated by the same criteria. Since states with smaller λ difference tend to have better energy derivative distribution overlap, $\Delta A_{AB}$ can be computed as the sum of free energy differences between several neighboring states connecting two end states:

$$\Delta A_{AB} = \sum_{i=1}^{N-1} \Delta A_{i,i+1} \; . \qquad (5)$$

As shown later, with the increasing of number of copies N and neighboring energy derivative distribution overlap becoming larger, free energy convergence also becomes faster both due to more efficient conformational sampling and more confident free energy determination.

In biomolecular researches, chemical modifications are often made, such as mutating residues in protein binding sites or changing binding ligands; this may very possibly lead to local conformational rearrangements [1]. The present method is especially suitable for relative binding affinity calculations in these systems. To illustrate the present method, alchemical free energy simulation between amino acids asparagines (ASN) and leucine (LEU) is taken as the model system. As shown in reference 9, both asparagines and leucine have multiple conformations related to sidechain $\chi_1$ and $\chi_2$ angles and these conformations are separated by high energy barriers into various minimum regions. For the reason that global minima of ASN and LEU are located in different regions, correct free energy difference cannot be easily obtained by regular free energy simulation technique, such as TI. This can occur because global minimum structure of starting state is evolved to be one of local minimum structures of ending state with λ switching if no special treatment is applied.

In the setup of DT-HREM simulation (Figure 1a), ASN and LEU share the same amino acid backbone structure (neutrally blocked) with its internal potential corresponding to part of $U_E$ in Eq. 3. Scaled energy potential terms (Van der Waals, Coulomb and Torsional terms) determining ASN and LEU sidechain conformations correspond to $U_S^A(\overrightarrow{X})$ and $U_S^B(\overrightarrow{X'})$ in Eq. 3 respectively; unscaled energy potential terms (Bond and Angle terms) of amino acid sidechain portions are also part of $U_E$. So at one of two end points, we have one of amino acids' energy potentials together with harmonic energy terms of the other state's side chain. Because free energy change due to harmonic energy terms is separable from that caused by the rest of system [13], the contributions from the harmonic energy terms of the other state can be cleanly cancelled when computing the change of free energy difference in two environments, for instance, from gas phase to solution for solvation free energy difference of these two amino acids, or from solution to protein for their binding free energy difference. Unscaled environment energy terms in $U_E$ can keep certain structural integrity and facilitate the increase of exchange frequency at the end points ($\lambda = 0$ for $U_S^B(\vec{X})$ or $\lambda = 1$ for $U_S^A(\vec{X})$) by increasing energy derivative distribution overlap between neighboring state ensembles. This is one of identified advantages of HREM methods [10]. If applied in protein systems, solution part and most of protein can be treated by unscaled environment term $U_E$, and conformation-determining energy terms (such as Van der Waals, Coulomb and Torsional terms) at mutation site and its local neighborhood, which may undergo conformational change upon mutation, should be duplicated and scaled in the way described in Eq. 3. Choice of scaled portion in the setup of DT-HREM simulation is important to balance enhanced sampling efficiency and accuracy, which will be discussed later.

All the calculations were carried out in the customized version of BLOCK module in CHARMM program [14]. In the present work, most of simulations were run at $\lambda = 0$, 0.1, 0.2 0.45, 0.55, 0.8, 0.9, 1 with 8 CPUs. In comparison, one simulation was run at $\lambda = 0$, 0.4, 0.6, 1 with 4 CPUs. In all calculations, a time step of 1 fs was used. The mass of the solute hydrogen atoms was set to 10 amu. To control the temperature, we used Langevin dynamics with a friction coefficient of 60 ps$^{-1}$ acting on all nonhydrogen atoms and random forces corresponding to the target temperature of 300 K. For DT-HREM simulations, exchange frequency was set as one per 100 timestep. To obtain free energy difference between states $\Gamma(\lambda_i)$ and $\Gamma(\lambda_j)$, "universal" probability density function (UPDF) was utilized to fit work distribution functions $f(W_{ij})$ and $g(W_{ij})$. And then intersection of two curves (by Nanda & Woolf etc. in ref 5), which should be free energy difference of these two states based on Equation 2, was determined.

As discussed above, for each pair of scale parameters $\lambda_i$ and $\lambda_j$, forward work and backward work $W_{ij}$ is equal to $\Delta\lambda_{ij}$ times energy derivative $\left.\frac{\partial U}{\partial \lambda}\right|_{\lambda=\lambda_i}$ and $\left.\frac{\partial U}{\partial \lambda}\right|_{\lambda=\lambda_j}$ at two $\lambda$ runs. So the value at the intersection of two energy derivative distribution functions scaled by $\Delta\lambda_{ij}$ should be equal to free energy difference $\Delta A_{ij}$. Energy derivative distribution functions for $\lambda = 0, 0.1, 0.45, 0.9, 1$ from 8 CPU run are shown in Figure 2. Obviously, choosing only five $\lambda$ values is enough to guarantee good overlap in between neighboring distributions covering the $\lambda$ range from 0 to 1, which as discussed above is the criteria of both confident determination of free energy and efficient sampling. In replica exchange simulations, usually a short preliminary run is carried out to determine optima set of temperatures or $\lambda$ values. The purpose of the present work is to demonstrate how to efficiently compute free energy difference using DT-HREM; so $\lambda$ values have not been on-purpose optimized. As shown (green line) in Fig. 3, free energy results obtained by DT-HREM overlap histogramming method converge at 300 ps with the value –61.7 kcal/mol and are robustly stable afterwards. This value is quantitatively consistent with what is obtained (-61.7 kcal/mol) by the study of Boresch etc. using adaptive umbrella sampling TI approach with much longer simulation time (about 150 ns) [9]. As comparison (red line in Fig. 3), when eight regular molecular dynamics simulations on the same sets of $\lambda$-scaled potentials with no replica exchange were run, -59.7 kcal/mol was stably obtained after long time simulation. This value is consistent with one of incorrect values obtained in the study by Boresch etc. Apparently without the mechanism of replica exchange, sampling can not be efficient and one of the end states is trapped in local minimium. Theoretically, in the framework of DT-HREM, increasing work distribution overlap between neighboring runs brings nearly 100% synergy to two traditionally orthogonal problems, work distribution intersection point determination and configurational sampling in free energy simulations. Usually, improving efficiency in either aspect requires a great deal of efforts. When we reduce number of replicas, as shown (cyan line) in Fig. 3, 4-CPU run has much slower free energy convergence with free energy value approaching -60.5 kcal/mol after 1 ns simulation. This is apparently due to the reduction of neighboring runs' energy derivative distribution overlap with the number of replicas decreased.

Figure 4. shows the conformation distributions of ASN and LEU along $\chi_1$ and $\chi_2$ at eight different $\lambda$ runs. At high $\lambda$ region, ASN conformations along both $\chi_1$ and $\chi_2$ directions are distinctly separate by high energy barriers. Barrier crossing is facilitated by exchange mechanism of DT-HREM simulation with low $\lambda$ region conformations, which are more equally distributed. In a reverse way, at low $\lambda$ region, LEU conformations along both $\chi_1$ and $\chi_2$ directions are distinctly separate by high energy barriers and barrier crossing is facilitated by exchange mechanism of HREM simulation with high $\lambda$ region conformations.

The illustration in Figure 4 also indicates the importance of choosing energy terms to be scaled; energy terms (Van der Waals, electrostatic, torsional terms) determining both $\chi_1$ and $\chi_2$ related conformations in the present model system should be scaled by $\lambda$ or 1-$\lambda$ in order to explore energy landscape efficiently. Theoretically, all the energy terms can be scaled in DT-HREM method. However, increasing the number of scaled energy terms will likely decrease the overlap of work distribution function between neighboring $\lambda$ runs and so lower free energy simulation efficiency; especially at high effective temperature regime such as high $\lambda$ region for ASN and lower $\lambda$ region for LEU, structural integrity will be more likely to be destroyed when too many energy terms are scaled. For instance, in the present model, even if extra angle terms related to $\chi_1$ and $\chi_2$ angles are scaled in $U_S^A(\vec{X})$ and $U_S^B(\vec{X'})$, free energy calculation will converge much more slowly as shown by blue line in Figure 3. On the other hand, if we duplicate dual-topology copies at $C_\beta$ atom as shown in Figure 1b, although exchange frequency is increased, free energy convergence is slower than when energy terms related to both of angles are scaled, as shown in magenta line of Figure 3, because conformation along $\chi_1$ can not be explored efficiently.

As discussed above, based on the hybrid form of regular HREM simulation and classical form of "alchemical" path in free energy simulation [15], DT-HREM is developed by the realization that in DT-HREM, both replica exchange efficiency and work distribution intersection determination confidence rely on the same criteria: energy derivative distribution overlap between two neighboring states labeled by two λ values. By inserting more intermediate states between starting state and targeting state, two key problems in free energy simulation, especially in rough energy landscape, can be solved simultaneously. We summarize all the results obtained on free energy difference (ASN→LEU) calculations using different setups in Figure 3. In terms of its efficiency and reliability, the comparison in Figure 3 clearly demonstrates the power of the presented "DT-HREM" overlap histogramming method, especially in the systems with rough energy landscapes. Due to the nature of scaling local energy terms in DT-HREM method rather than heating up the whole system unselectively like in temperature replica exchange method, this technique also makes free energy simulations of local events in large systems such as ligand binding and chemical mutation robust. Currently, we are applying this method in broad range of drug binding and protein-protein binding problems. Due to its efficiency, accuracy and robustness, we believe, this method will contribute to computational biophysical applications a great deal.

Support by FSU research foundation and FSU CRC is gratefully acknowledged.

*Correspondence: yang@sb.fsu.edu (W.Y.); berg@scs.fsu.edu (B.A.B)

**Captions**

Figure 1. Setups for DT-HREM simulations on the "alchemical" transformation from Leucine to Asparagine. a) Leucine and Asparagine share neutrally blocked amino acid backbone; b) Leucine and Asparagine share both amino acid backbone and $C_\alpha$ atom, in which $\chi_1$ can not be sampled efficiently.

Figure 2. Energy derivative distributions for various ensembles on different $\lambda$ scaled potentials ($\lambda = 0, 0.1, 0.45, 0.9, 1$). Red line shows the result fitted by "universal" probability density function (UPDF).

Figure 3. Time evolution of computed free energy difference between Leucine and Asparagine with various DT-HREM overlap histogramming setups. If there is no special note, simulations were run with 8-CPU setup Red line: overlap histogramming result with no replica exchange allowed; Cyan line: DT-HREM overlap histogramming result with 4-CPU run; blue line: DT-HREM overlap histogramming result with additional angle terms scaled; magenta line: DT-HREM overlap hisogramming result with $\chi_1$ bond shared by two amino acids.

Figure 4. Conformation distributions of Leucine and Asparagine along $\chi_1$ and $\chi_2$ angles in various ensembles on different $\lambda$ scaled potentials. These conformational distributions were obtained in DT-HREM simulations.

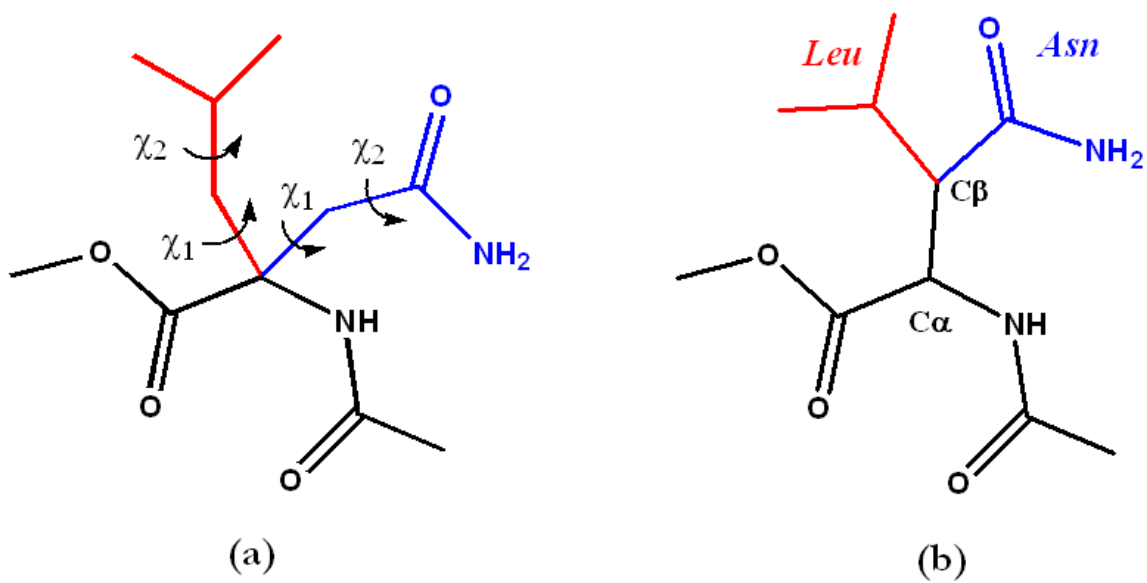

**(a)**

**(b)**

Figure 1.

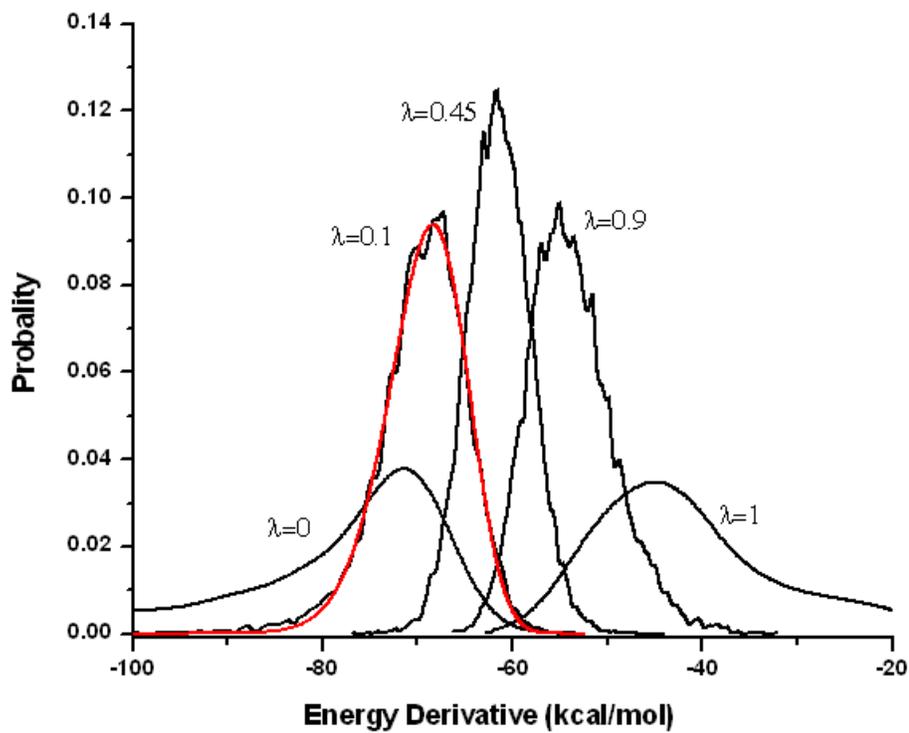

Figure 2.

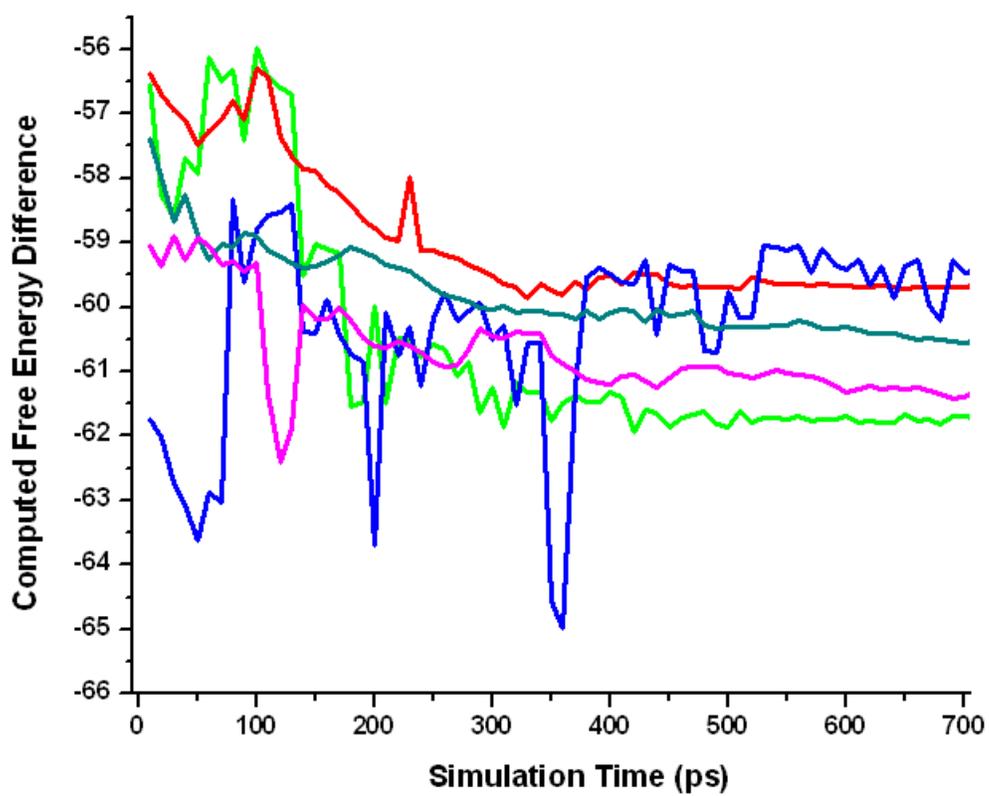

**Figure 3.**

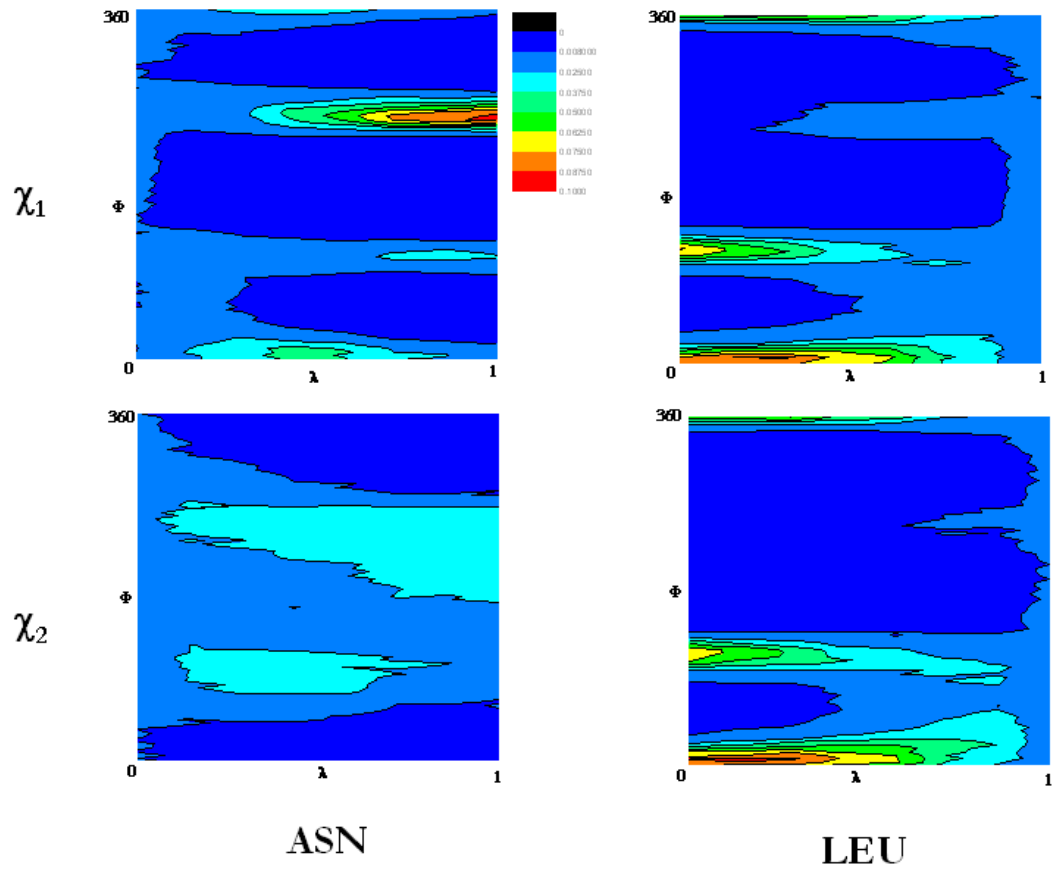

$\chi_1$

$\chi_2$

ASN LEU

Figure 4.